\documentclass[aps,prl,twocolumn,groupedaddress]{revtex4}
\usepackage{epsfig}
\usepackage{amsmath}
\usepackage{amsfonts}
\usepackage{amssymb}

\begin{document}
\newcommand{\eqbeg}{\begin{equation}}
\newcommand{\eqend}{\end{equation}}
\newcommand{\shat}{\hat{S}}
\newcommand{\ddt}{\frac{d}{dt}}
\newcommand{\thalf}{\mbox{T}_{1/2}}

\title{Experimental Study of Acoustic Ultra-High-Energy Neutrino Detection}

\author{J.~Vandenbroucke}\altaffiliation[Now at ]{the University of California, Berkeley}
\author{G.~Gratta} 
\author{N.~Lehtinen}\altaffiliation[Now at ]{the University of Hawaii, Manoa}

\affiliation{
	Physics Department, Stanford University, Stanford CA 94305 \\
 }

\date{\today}

\begin{abstract}
An existing array of underwater, large-bandwidth acoustic sensors has been used to study
the detection of ultra-high-energy neutrinos in cosmic rays.    Acoustic data from a 
subset of 7 hydrophones located at a depth of $\sim 1600$~m have been acquired for a total live 
time of 195 days. For the first time, a large sample of acoustic background events has 
been studied for the purpose of extracting signals from super-EeV showers.  As a test 
of the technique, an upper limit for the flux of ultra-high-energy neutrinos is presented 
along with considerations relevant to the design of an acoustic array optimized for 
neutrino detection.

\end{abstract}
\maketitle

\section{Introduction}

The detection of a neutrino component in cosmic rays above $10^{19}$~eV is a central goal of particle astro-physics.  It is
generally believed~\citep{G,ZK} that the flux of cosmic ray protons should drop sharply (``GZK'' cutoff) between $10^{19}$
and $10^{20}$~eV due to pion photo-production on the microwave background.  Recent measurements using the AGASA surface
array~\citep{AGASA} and the HiRes air fluorescence telescope~\citep{HiRes} seem to disagree on the existence of such a cutoff,
although it has been pointed out~\citep{BahcallWaxman} that they may actually be consistent with one another. Neutrinos are
likely to be an important component of the particle flux at ultra-high energies.  If the GZK mechanism is indeed present,
the flux of ultra-high energy protons must be accompanied by ultra-high energy neutrinos (``GZK'' neutrinos) from pion
decay.  Their clear detection would provide crucial input to the field.

Detection of the very small flux of cosmic rays at these extreme energies requires detectors of unprecedented size.  
Therefore only detectors employing naturally occurring media are practical.   The parallel development of several
different techniques, some of which would ideally employ the same medium, is therefore crucial to constraining the
measurements and providing sufficiently accurate and redundant information in order to fully understand cosmic rays at the
highest energies. Indeed, in the last few years, great progress has been made in the use of radio signals for the detection
of protons and neutrinos using a variety of natural targets, including lunar regolith~\citep{GLUE} and ice with
ground-~\citep{RICE} and space-based~\citep{FORTE} antennas.

In this paper we describe the first experimental results obtained with the acoustic technique in ocean water.  The
possibility of acoustic detection of high-energy particles was first discussed in 1957~\citep{Askaryan,AskDolg}.  Following this
work, an extensive theoretical analysis of the processes relevant to signal formation was performed~\citep{Learned}, and
the general phenomenon was confirmed with an accelerator test beam~\citep{Sulak}.  The first description of the properties of
a practical, large array for ultra-high-energy neutrino detection was given in~\citep{Lehtinen}. The dominant sound
production mechanism, for non-zero water expansivity, consists of water heating in the region where a shower develops
(instantaneously, from the point of view of acoustics), followed by expansion, producing a pressure wave.  Only primaries
that penetrate the atmosphere without interacting and then shower in the water are capable of producing detectable acoustic
pulses, providing automatic particle identification within a broad range of energies.  Acoustic sensor arrays can be
substantially more sparse than optical Cherenkov arrays in water or ice, because of the large ($\approx 1$~km) attenuation
length of sound in water at the frequencies ($\approx 10$~kHz) where, for these events, most of the energy lies. However,
as stressed in~\citep{Lehtinen}, the elongated shape of the showers produces acoustic interference that results in most of the
sound emitted in a pancake-shaped volume with few degrees opening angle. This effect is important for the optimization of a
detector since the optimal density of hydrophones is constrained by the ability to intercept these relatively narrow
pancakes.  Finally the acoustic technique lends itself well to a calorimetric measurement of the total energy and
would complement other technologies well.

\section{The Detector}

\begin{figure}[thb!!!!!!!!!]
\begin{center}
\mbox{\epsfig{file=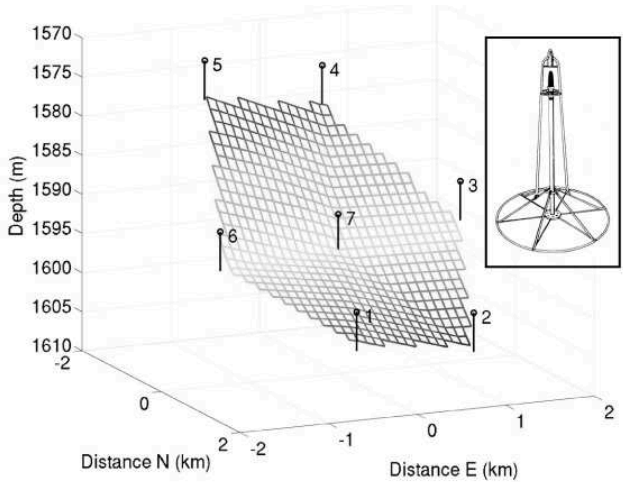,width=8.5cm}}
\caption{
Schematic view of the subset of hydrophones used for this work.    The hydrophone 
depths are known precisely while the sea floor bathymetry between phones is estimated with a 
linear interpolation.   The inset shows the 4.5~m tall hydrophone
deployment and support structure.   The bail used to deploy the phone folds and lies 
on the ocean floor during operation.} 
\label{fig:3D_detector}
\end{center}
\end{figure}

SAUND (Study of Acoustic Ultra-high energy Neutrino Detection\footnote{http://hep.stanford.edu/neutrino/SAUND}) employs a
large ($\sim250$~km$^2$) hydrophone array in the U.S. Navy Atlantic Undersea Test And Evaluation Center 
(AUTEC)\footnote{http://www.npt.nuwc.navy.mil/autec/}.  The array is located in the 
Tongue of the Ocean, a deep tract of sea in the Bahama islands
at approximately $24^{\circ} 30^{\prime} N$ and $77^{\circ} 40^{\prime} W$.  A detailed description of the entire array is
given in~\citep{Lehtinen}.  For the present work, we use a subset of seven hydrophones from the large array, arranged in a
hexagonal pattern as shown in Figure~\ref{fig:3D_detector}. The hydrophones are mounted on 4.5~m booms standing vertically
on the ocean floor.  The horizontal spacing between central and peripheral hydrophones is between 1.50~km and 1.57~km. The
ocean floor is fairly flat over the entire array region, with our phones located at depths between 1570~m and 1600~m.  The
hydrophones and their underwater preamplifiers have been in stable operation for naval exercises since 1969.

Analog signals are amplified at the hydrophones and fed through an on-shore amplifier stage to a digitizer 
card\footnote{National Instruments Model PCI MIO 16E1} that continuously samples each of the seven channels at 179~ksamples/s.  The time series 
in each channel is then fed to a digital matched filter implemented on a 1.7~MHz Pentium-4 workstation.

Ultra-high energy neutrinos interact with matter by deep inelastic scattering on quarks inside nuclei.  With increasing
energy, the interaction cross-section~\citep{Kwi}, and hence the probability of a neutrino interacting within the ocean water
(which is thinner than the $\sim100$~km interaction length at 10$^{21}$ eV), increases.  After the primary interaction, the
neutrino energy is distributed between a quark and a lepton.  On average, the lepton acquires $\sim$80\% of the
energy~\citep{Gandhi}.  The remaining energy is dumped into the water as a hadronic shower aligned with the direction of the
primary neutrino.  The water is heated locally, causing it to expand and emit an acoustic pulse propagating perpendicular to
the shower axis.

\begin{figure}[bht!!!!!!!!!!!!!!!!!!]
\begin{center}
\mbox{\epsfig{file=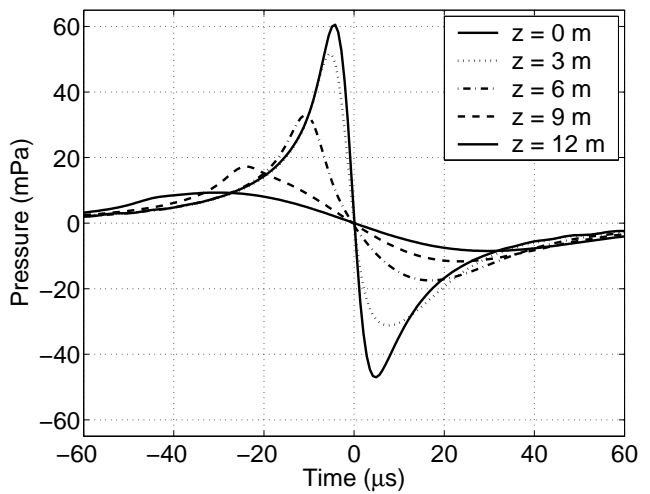,width=8.5cm}}
\caption{
Simulated acoustic signals due to a hadronic shower of energy $10^{20}$~eV,
corresponding to an initial neutrino energy of $5 \times 10^{20}$~eV.  Signals are 
shown at 1050~m distance perpendicular to the neutrino path.  The five different 
curves refer to longitudinal distances of 0, 3, 6, 9, and 12~m forward from the 
shower maximum.
} \label{fig:neutrino_signals}
\end{center}
\end{figure}

While a muon or tau neutrino generates only a hadronic shower, an electron neutrino also generates an electromagnetic shower, which is superimposed on the
hadronic one. However, due to the Landau-Pomeranchuk-Migdal (LPM) effect~\citep{LP,Migdal1,Migdal2}, the electromagnetic shower is elongated and has an irregular
structure, producing weaker acoustic signals with irregular geometries.  Here we only consider the hadronic showers, which exhibit the same structure for all
kinds of neutrinos and are simulated with a simple parametrization as modeled in~\citep{Alvarez}.  This model includes $\pi_0$ interactions and the LPM
effect for the electromagnetic components of the hadronic shower.  Once the energy deposition has been determined using this shower model, the resulting acoustic
pulse shape is simulated, following~\citep{Learned,Lehtinen}, at arbitrary positions ($r, \theta$)  with respect to the shower.  As shown in
Figure~\ref{fig:neutrino_signals} for a fixed radial distance from the shower, the bipolar pulse is tallest and narrowest in a plane perpendicular to the shower
axis, at the longitudinal location of the shower maximum.\footnote{We note here that an erroneous value for the volume expansivity of seawater was used
in~\citep{Lehtinen}.  A value of $1.2 \times 10^{-3}$ was used instead of the correct value of $~0.15 \times 10^{-3}$.  This resulted in a simulated signal
amplitude larger than the correct value by a factor of 8.  This error is corrected in the present work.} The pulse becomes shorter and wider at greater
longitudinal distances from the maximum.  The general features of the pulse shape match the data from~\citep{Sulak}.

Triggering is achieved in SAUND with a digital matched filter that searches for signals matching the expected neutrino
signal, appropriately weighting frequency components according to their signal-to-noise ratio~\citep{Filters}.  The filter
response function is obtained by transforming to the frequency domain the expected signal shape, which is approximated in
the time domain by the analytical form $S(t) \propto - (t/\tau)e^{-t^2/{2 \tau^2}}$.  The characteristic time of the signal
depends on position $(r, \theta)$ relative to the shower (see Figure~\ref{fig:neutrino_signals}). A characteristic time of
$\tau = 10$~$\mu$s, corresponding to the width of the largest expected signals, is chosen for the present work. To obtain
the filter transfer function, the expected signal in the frequency domain, $\tilde{S}(f)$, is divided by the noise spectrum,
assumed to be of the Knudsen type~\citep{Knudsen}: $\tilde{N}(f)\propto f^{-\alpha}$.  The expected $\alpha=1.7$ is
approximated with $\alpha=2$ for analytical convenience.  The transfer function $\tilde{H} \propto
\tilde{S}(f)/\tilde{N}(f)$ is then transformed to the time domain to obtain the response function, \begin{equation} H(t) =\left[
\left(\frac{t}{\tau}\right)^3-3 \left(\frac{t}{\tau}\right) \right] e^{-t^2/{2\tau^2}}.  \end{equation}

Each channel is continuously filtered and the filter output 
is compared to a threshold.  The filter is implemented
with a discretized version of the above response function, sampled at the same frequency as the
pressure waveforms (179 kHz).  12 samples ranging from $t=-30.8~\mu$s to 
$t=30.8~\mu$s are used.  Beyond this range, the response function is below 13\% of its peak amplitude.    
When the threshold is exceeded, a trigger occurs and event data are recorded.

Signals from the highest-energy neutrinos considered here are expected to saturate the
digitizer, and $\sim10\%$ of waveforms in the dataset are saturated.  By simulating the
effects of saturation on the digital filter and on subsequent offline cuts, we found that
saturation does not pose a problem to the present analysis.  This is because the cuts rely
primarily on frequency and timing information, not on amplitude.

\begin{figure}[bt!!!!!!!!!!!!!!!!!!!!!!!!!!!!!!!!!]
\begin{center}
\mbox{\epsfig{file=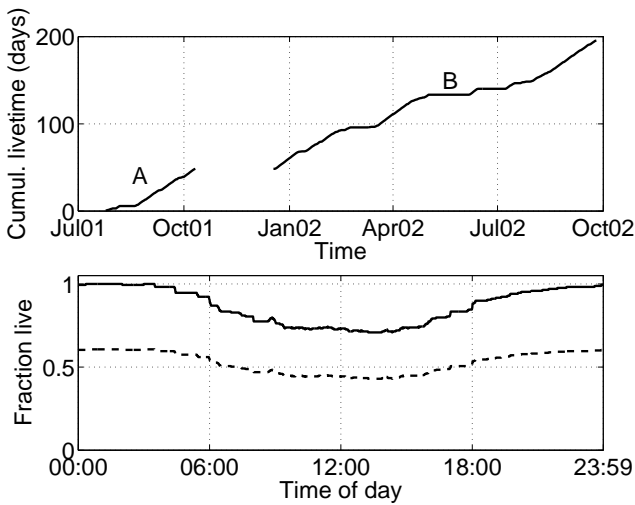,width=8.5cm}}
\caption{
Live time for SAUND. In the top panel, we show the data accumulation as a function of time. A test run
(Run A, 48 days), not used in the present analysis, was followed by a system upgrade and a longer period
of stable running (Run B, 147 days).  In the bottom panel, the solid curve gives the fraction live
as a function of time of day, for days during which any SAUND Run B data were collected.  The dashed 
curve represents the data used for the analysis reported here, and it includes down time from bad weather 
and hardware failures.}
\label{fig:livetime}
\end{center}
\end{figure}

By arrangement with the U.S. Navy, the SAUND data acquisition system (DAQ) is connected to the
hydrophone array only when the array is not in use for Navy exercises.  The live time achieved
under this agreement is shown in Figure~\ref{fig:livetime}.  As shown in the top panel, a test 
run (Run A) was used to commission the system and is not used for the present analysis. 
The system was upgraded from Oct. 2001 to Dec. 2001 and run for an integrated 147 days live 
with stable conditions (Run B).  Only this stable run is used for the analysis presented here.  
The solid curve in the bottom panel shows the live time achieved at different times of day.    
The live time is close to 100\% at night, when no Navy tests are conducted, and it drops to
about 70\% in the middle of the day.    The average over a day is $86\%$.    In addition,
an overall live time reduction due to weather conditions and hardware failures resulted in 
the dashed line in the bottom panel of Figure~\ref{fig:livetime}.   

\begin{figure}[htb!!]
\begin{center}
\mbox{\epsfig{file=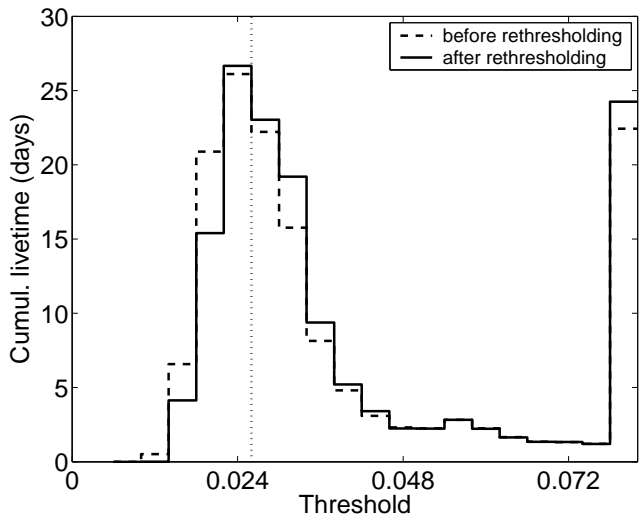,width=8.5cm}}
\caption{
Integrated live time at each discrete threshold.  The distribution has a long tail extending 
beyond the range shown here; this tail has been added to the last bin of the figure.  The 
threshold step size is 0.004.  The four values from 0.012 
to 0.024 (left of the vertical line) are considered ``quiet'' and used for the present analysis.  Including 
data taken at larger thresholds would increase the background while negligibly increasing neutrino 
exposure.  The dashed and solid histograms represent the same quantity before and after, respectively, the 
offline re-thresholding described in the text.
} \label{fig:quiet_livetime}
\end{center}
\end{figure}

Because the noise environment is volatile, an adaptive trigger threshold is used.  Every minute the threshold is raised or
lowered based on whether the number of events in the previous minute was above or below the target rate of 60~events/minute.  
Events are accumulated in memory each minute and written to disk, along with data relevant to the entire minute, at the end
of the minute. The distribution of thresholds in Run B is shown in Figure~\ref{fig:quiet_livetime}.  To simplify online
processing and offline event reconstruction, the threshold value is constrained to discrete values (integer multiples of
0.004, in units of digital filter output).  For the purpose of neutrino analysis, only thresholds between 0.012 and 0.024
are used.  These ``quiet'' periods account for 37\% of Run B.

In total, Run B contains 20~M triggers (250~GB) corresponding to
$\sim$1.7~GB/day.  Every few weeks, data are transferred to an external hard drive that is
shipped to Stanford for offline analysis.  Nine-tenths of the triggers are captured for 1~ms;
one-tenth are captured for 10~ms in order to record and study reflections from the ocean
bottom.

The frequency response of the hydrophone/amplifiers chain is flat to better than 
8~dB between 7.5~kHz and 50~kHz.  A sharp cutoff of 100~dB/octave below 7.5~kHz is due to
an analog high pass filter in the system, while a somewhat slower roll-off, due to
the sensor itself, occurs above 50~kHz.    The low frequency cutoff, while not ideal for 
neutrino detection, is not a significant hindrance.

The phase response of the hydrophone/amplifiers chain is difficult to measure and at present is not well known.  In
order to gain some quantitative understanding of its possible effects on the trigger, the efficiency of the filter was
calculated for the pulse shapes in Figure~\ref{fig:neutrino_signals} and for their first and second time derivatives, after
applying a 9th-order Butterworth bandpass filter to mimic the AUTEC frequency response.  For each of these phase
responses, efficiency was calculated as a function of energy.  The resulting trigger thresholds have a spread of a factor
of two for any given energy.   Conservatively, the highest of these thresholds is assumed in the present analysis.

In addition to the matched filter described above, an online veto is applied to remove a
specific kind of noise that is time-correlated in all channels.  This noise consists of 
simultaneous sinusoids in all 7 channels, occurring 
in bursts with a 60~Hz repetition rate.  While the origin of these signals is unknown, it is clear 
that they originate from electrical pickup, because the sound velocity does not allow for 
simultaneous, correlated acoustic events in all phones.
The veto removes this noise by rejecting events with large 7-channel pairwise correlation,
\begin{equation}
C = \sum_{i<j} \sum_{t} p_i(t) p_j(t),
\end{equation}
where the time sum is taken over discrete samples and $p_i$ is the pressure time series at 
phone $i$, normalized by mean absolute amplitude: $p_i = P_i / \langle |P_i| \rangle$ where $P_i$ is the 
absolute pressure and $\langle |P_i| \rangle$ is its average over the captured waveform.  
Events with $C > 500$ are removed.

\begin{figure}[htb!!] 
\begin{center} 
\mbox{\epsfig{file=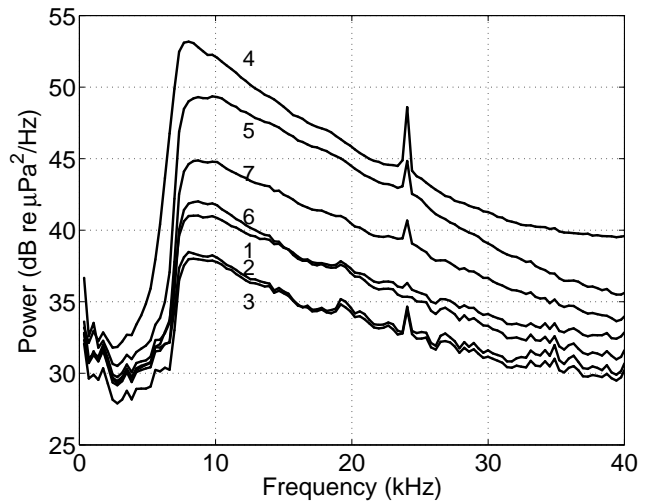,width=8.5cm}}
\caption{A typical average noise spectrum as measured by SAUND in Run B.  
The plot shows the spectrum averaged over 10 hours from midnight to 10:00 AM on Dec. 22,
2001.  The range of noise levels is mainly due to variations in amplifier gains. Note 
the 7.5~kHz high-pass filter and the $\sim 24$~kHz peak from a transmitter used by the Navy.  The sea 
state noise dominates these spectra.}
\label{fig:noise_spectrum} 
\end{center} 
\end{figure}

Every 10 minutes, time series for all channels are collected for 100~ms in forced mode. These data are used for offline
noise analysis and for simulating the noise conditions in our efficiency calculation.  Every minute the noise spectrum is
calculated from a 100~ms-long time series.  A typical noise spectrum is given in Figure~\ref{fig:noise_spectrum}.

SAUND Run B provides the largest data set ever collected for the purpose of studying acoustic techniques in UHE particle
detection.  During the run it has been our strategy to collect data with consistent and stable conditions for an extended
period of time.  A number of improvements in trigger efficiency, stability and uniformity have been developed from Run B and
will be implemented in future runs.

\section{Energy and Position Calibration}

Small implosions are useful calibration sources for underwater acoustics.  Household light bulbs, which implode at a
particular failure pressure while sinking, releasing energies of order 100~J~\footnote{
			For the most common household (``A19'' sized)
                             incandescence bulbs ($\sim 6.7$~cm diameter);
                             no signal was detected from miniaturized bulbs ($\sim 1$~cm diameter),
                             possibly because no implosion occurs.} , are a particularly convenient source~\citep{Heard}. A calibration run of this type was
performed on the morning of July 30, 2001 (at the beginning of Run A). 

The small boat used for the calibration was positioned over the central hydrophone using a
handheld Global Positioning System (GPS)  receiver\footnote{Magellan, Model 310}.  The boat engine was then turned off and several light bulbs were sunk as the
vessel drifted for about thirty minutes.  Two more GPS fixes were acquired during the calibration.

\begin{figure}[htb!!] 
\begin{center} 
\mbox{\epsfig{file=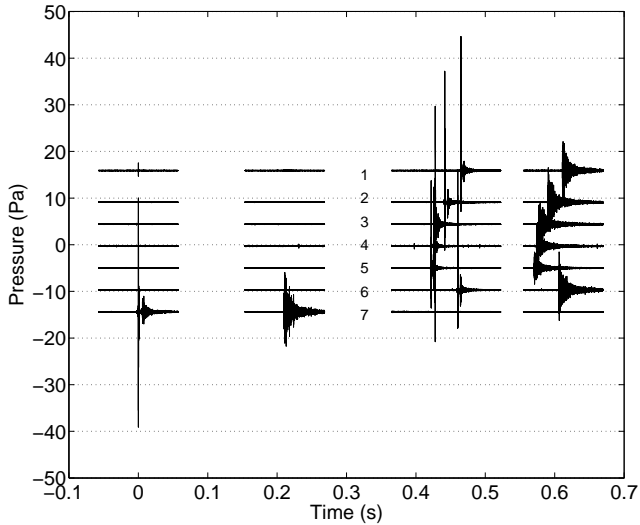,width=8.5cm}} 
\caption{Signals from a light bulb implosion.  
Time series from all seven hydrophones are shown.   A direct signal at the central 
hydrophone is immediately followed by a bottom reflection and, much later, by the signal 
from the reflection off the ocean surface.  Similar sets of 3 signals appear, at a later 
time, on the surrounding six hydrophones.  The signals in different channels are 
displaced vertically for clarity.   The small signal in phone 1 at $t =$ 0 is electrical cross-talk due to the large signal 
in phone 7.
}\label{fig:bulb_all_phones} 
\end{center} 
\end{figure}

\begin{figure}[htb!!] 
\begin{center}
\mbox{\epsfig{file=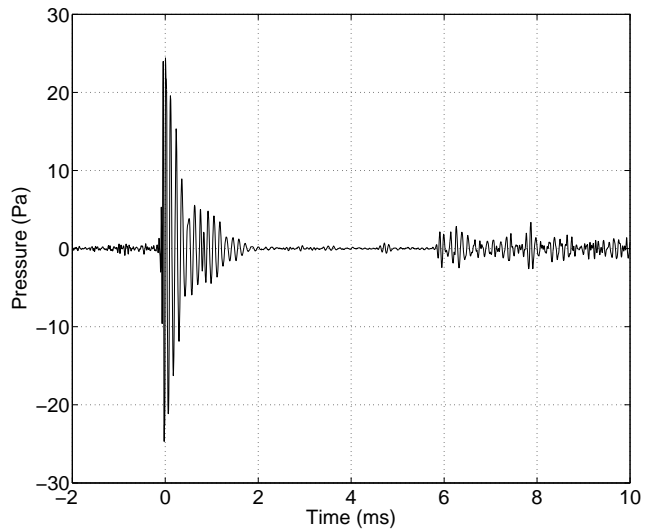,width=8.5cm}}
\caption{Detail of the light bulb implosion signal at the central hydrophone.  Both the direct signal and, after
a 6 ms delay, the bottom reflection, are evident.  The oscillations in the envelope of the direct signal are due 
to a combination of the bubble rebound from the implosion~\citep{Heard} and the response of the hydrophone, 
cable and amplifiers.} 
\label{fig:bulb_one_phone} 
\end{center} 
\end{figure}

The waveforms recorded for one of the implosion events are shown in
Figure~\ref{fig:bulb_all_phones}.  The first pulse corresponds to the
direct acoustic signal arriving at the central hydrophone.  The pulse
around 0.2~s corresponds to the arrival at the hydrophone of the signal
reflected by the surface of the ocean.  Direct pulses in all other six
hydrophones appear between 0.4~s and 0.5~s, followed, in turn, by the
respective surface reflections.  In Figure~\ref{fig:bulb_one_phone} the
waveform for the central hydrophone is shown on an expanded time scale,
clearly showing the bottom reflection around 6~ms.

Also shown in Figure~\ref{fig:bulb_one_phone} is the structure of the direct pulse.  The signal is expected to be a
damped sinusoid, generated by the ``breathing"  oscillations of the gas
bubble produced by the bulb.  In our case, the primary resonant frequency of 
the bubble, at a depth of 100-200~m depth, is predicted to be 
700-1300~Hz~\citep{Heard}.  The signal shown in Figure~\ref{fig:bulb_one_phone} 
shows a second peak at about 1~ms, confirming this prediction.  The measured waveform is consistent with 
the predicted waveform after a 7.5~kHz high-pass filter.

A time-difference of arrival (TDOA)~\citep{TDOA} analysis can be used to reconstruct the 3D location 
of acoustic events.  
This method uses only timing information.  In a homogeneous medium, the difference in arrival times 
between each pair of
hydrophones constrains the source location to a hyperboloid.  Four hydrophones (three independent 
time differences) are
necessary to determine three hyperboloids, the intersection of which can be found analytically and 
generally results in two
points.  For our sea-floor, planar array, the two solutions are usually symmetric about the detector 
plane, and the solution
below the sea floor can be discarded.

\begin{figure}[tb!!!!!!!!!!!!!!!!!!!!!!!!!!!!!!!!!]
\begin{center}
\mbox{\epsfig{file=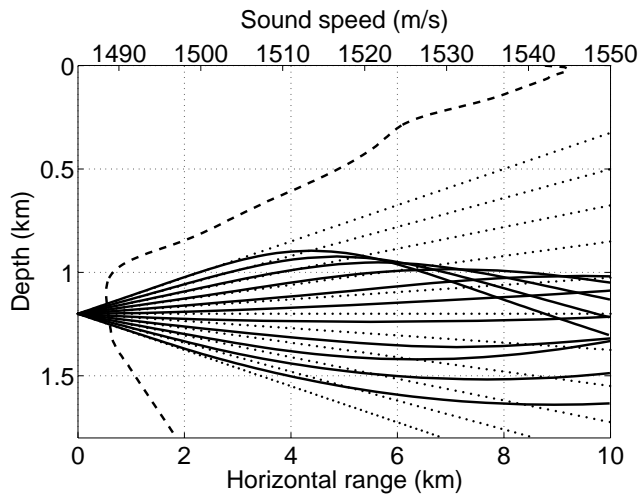,width=8.5cm}}
\caption{Example of ray trace showing the effects of refraction.  11 rays are 
emitted from a source at 1200 m depth, from $5^{\circ}$ below to $5^{\circ}$ above the
horizontal.  Refracted (solid) and unrefracted (dotted) rays are plotted together
for comparison.  For reference, the sound speed as a function of depth
(``sound velocity profile'', SVP), used to calculate the ray trace, is also shown
(dashed).
 } \label{fig:ray_trace_svp}
\end{center}
\end{figure}

While an analytical TDOA solution is possible in homogeneous media, the ocean is not homogeneous 
but is a layered medium.  At 
AUTEC, the sound velocity varies with depth as shown in Figure~\ref{fig:ray_trace_svp}. 
This results in sound ray refraction as illustrated in the same Figure.  
Over the fiducial volume of the SAUND detector refraction significantly affects position 
reconstruction.  The effect warps the TDOA hyperboloids, making an analytical solution impossible.
A numerical solution is sought by minimizing 
the metric \begin{equation} m(\vec{s}) = \frac{1}{N-1}\sum_{i<j} \left[t_{ij}^{meas} -
t_{ij}(\vec{s}) \right]^2 \end{equation} 
with respect to $\vec{s}$, the test location for the acoustic source.  
Here $t_{ij}^{meas}$ is the measured TDOA between phones $i$ and $j$ 
($t_{ij}^{meas} = t_i^{meas} - t_j^{meas}$) and $t_{ij}(\vec{s})$ is 
the theoretical difference in arrival times for a source at 
$\vec{s}$ ($t_{ij}(\vec{s}) = t_i(\vec{s})  - t_j(\vec{s})$).  
For $N$ phones recording a signal, the sum occurs over the $N-1$
independent pairs of arrival time differences.

The theoretical travel time from the test source location to phone $i$,
$t_i(\vec{s})$, is calculated using a ray trace algorithm that divides the
water into layers in each of which the sound speed is a linear function of
depth.  The ray follows the arc of a circle within each layer~\citep{Boyles}.  
Practically, the ray trace is too time intensive to be performed for each
test source location.  For each phone, a table of travel times from points on
a grid spanning the detector volume to the phone is built.  The minimization
begins at the best grid point, and the grid is linearly interpolated to find
the off-grid source location.  The grid is built once, and the interpolation
then occurs for each event reconstruction.  While the local sound velocity profile 
(SVP) is measured almost daily by the Navy and is available to us, it was found that
seasonal SVP variations introduce errors smaller than those due to the
uncertainty in the hydrophone location.   Hence a single table built from an
average SVP is used for all reconstructions.  The position reconstruction
algorithm takes a few seconds to run, for each localization, on a 1.6~GHz Athlon CPU (offline computer).

Because the light bulb implosions are far above noise, their TDOA localization is performed
using all 7 direct signals (6 independent time differences), rather than 5 as used for events triggered
in neutrino mode. The reconstructed depth of bulb failure, $D$, is then used to estimate the energy
released in the implosion, $P_{\rm amb}V$.   Here $P_{\rm amb} = \rho g D$ is the ambient pressure at the
reconstructed implosion depth, $\rho$ is the approximate density of sea water ($\rho = 1$~g/cm$^3$), $g$
is the acceleration due to gravity, and $V\sim 150$~cm$^3$ is the volume of the light bulb. 
From our position reconstruction $P_{\rm amb}$ is found to be of order 1000~kPa, while the internal
pressure of the bulbs is believed to be slightly below one atmosphere (70-90~kPa~\citep{Heard}).  The
total acoustic energy detected at phone $i$, $E_{\rm det}(i)$, can be calculated independently
assuming spherical emission~\citep{ElmoreHeald}:  
\begin{equation} 
E_{\rm det}(i) = \frac{4\pi r^2}{\rho c} \int P^2(i,t) dt, 
\end{equation} 
where $P(i,t)$ is the recorded
pressure time series at phone $i$, and $r$ is the distance from phone to reconstructed implosion
location.  Lensing effects due to the changing sound velocity as a function of depth were studied and
found to be negligible for these source locations.

\begin{figure}[tb!!!!!!!!!!!!!!!!!!!!!!!!!!!!!!!!!!!!!]
\begin{center}
\mbox{\epsfig{file=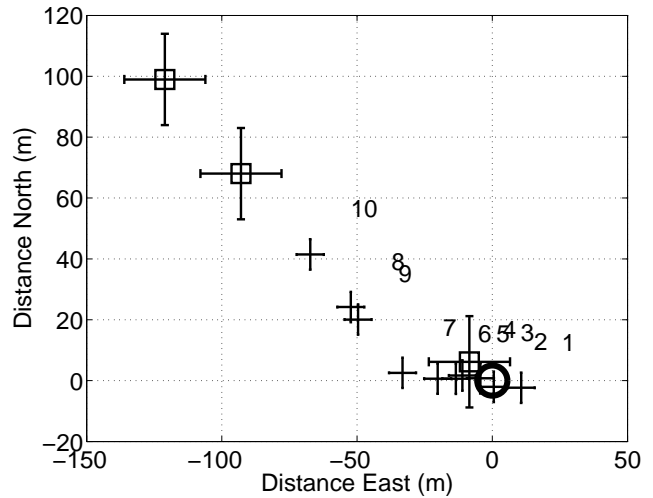,width=8.5cm}}
\caption{The trail of implosions of bulbs dropped from a drifting boat, as
reconstructed from time-differences of arrival.  The reconstructed positions are
indicated by crosses, with bulb labels offset for clarity.  The boat's motor was cut
after starting approximately above the central hydrophone (marked by a circle).  The
larger crosses represent the boat positions measured at three times with the GPS 
device.    Bulbs are numbered by the order in which they were dropped.  
The bulb drop spanned one half hour.
} \label{fig:bulb_trail}
\end{center}
\end{figure}

The horizontal reconstructed positions of the light bulb implosions are shown in 
Figure~\ref{fig:bulb_trail}, along with the three GPS fixes of the boat from which they
were dropped.  The implosions are sequentially numbered in the order they occurred.  
The GPS points, shown with the $\pm 15$~m errors quoted by the specifications of the 
instrument, are in good agreement with the TDOA reconstruction.  The extent of the drift is 
consistent with a current of about 5~cm/s (0.1 knot).

The results of depth and energy reconstruction for the implosions are shown in 
Table~\ref{tab:implosions}.  The uncertainty of position reconstruction near the 
surface directly above the center of the array (the optimal location for source 
localization) is 5~m.  The implosion pressures $P_{\rm amb}$ are consistent with the 
literature~\citep{Heard}.  The efficiency of conversion from $P_{\rm amb}V$ to acoustic
energy is expected to be about 4\%~\citep{Heard}.  In addition the acoustic frequency 
spectrum from the implosion is known to be peaked around the air bubble resonant 
frequency of $\sim 1$~kHz, well below AUTEC's low frequency 
cutoff~\citep{bulb1,bulb2,bulb3}, so that only about one-half of the acoustic energy 
lies within AUTEC's bandwidth.  The expected energy, taking into account both the
4\% yield and the bandwidth mismatch, is shown in column 5 of Table~\ref{tab:implosions}.
From energy reconstructed using recorded amplitudes (column 6), we find that the yield 
is in practice somewhat lower and it fluctuates substantially from one event to another.    
Both these effects
are possibly due~\citep{Heard} to the fact that the bulbs are allowed to implode at 
failure pressure rather than being intentionally broken at lower ambient pressure.

\begin{table} 
\caption{\label{tab:implosions}
Parameters of the implosion calibration events.  $D$ is the reconstructed implosion depth and
$P_{\rm amb} = \rho g D$ is the ambient pressure at $D$.   Here $\rho\simeq 1$~g/cm$^3$ is
the density of sea water and $P_{\rm amb}V$ is the implosion energy, with $V =~150~\rm cm^3$ the 
bulb volume.  $E_{\rm theor}$ is the theoretical detectable
acoustic energy, assuming 4\% conversion of the energy $P_{\rm amb}V$, as appropriate for 
implosions triggered with a hammer device~\citep{Heard}.   $E_{\rm theor}$ includes the 
effects of absorption and SAUND's frequency response.   The somewhat lower yield
observed ($E_{\rm det}$) is consistent with the expectation of lower acoustic coupling
for implosions allowed to occur at failure pressure rather than forced at lower pressure with 
a hammer device~\citep{Heard}.
} 
\begin{tabular}{lllllll}
Bulb      &       $D$       &   $P_{\rm amb}$    & $P_{\rm amb}V$ & $E_{\rm theor}$ &    $E_{\rm det}$  \\
          &       (m)       &        (kPa )      &       (J)    &    (J)	&      (J)      \\
\hline
 1        &       170       &        1640        &        250   &  3.1		&	1.7    	\\
 2        &       110       &        1120        &        170   &  1.9		&	0.3    	\\
 3        &       150       &        1430        &        210   &  2.6		&	1.5    	\\
 4        &       170       &        1690        &        250   &  3.2		&	2.8    	\\
 5        &       130       &        1300        &        200   &  2.3		&	0.8    	\\
 6        &       110       &        1050        &        160   &  1.8		&	0.4    	\\
 7        &        90       &         900        &        140   &  1.5		&	0.1    	\\
 8        &       140       &        1380        &        210   &  2.5		&	1.9    	\\
 9        &       200       &        1930        &        290   &  3.8		&	1.9    	\\
 10       &       300       &        2930        &        440   &  6.7		&	1.8   	\\
\end{tabular}
\end{table}

\section{Data Reduction}

The online trigger system is designed to select impulsive bipolar signals while
rejecting most of the Gaussian noise.   Because of the novelty of the technique,
trigger conditions are chosen to be rather loose and substantial background is left 
to be rejected by off-line analysis.    Impulsive backgrounds to UHE showers are expected to
arise from a variety of human and non-human sources.    In particular a number of animals, 
from large cetaceans to small crustaceans, are known to produce high-frequency acoustic 
transient noises.  While the Tongue of the Ocean is relatively isolated from the open ocean 
and hence rather quiet, an important motivation for this first large-volume data-taking campaign 
is the exploration of such coherent backgrounds.

Offline data reduction is used to select events consistent with neutrino-induced showers while rejecting backgrounds in
SAUND Run B. The efficiency of each step of data reduction is estimated using simulated neutrino events to which, as
discussed above, bandpass filtering is applied to account for the known frequency response of the array.  As is also
described above, the unknown phase response of SAUND is addressed by repeating the efficiency analysis with the first and
second derivative of the simulated and filtered signal.  The worst case (lowest efficiency) is then assumed.  The overall 
efficiency of the data reduction cuts is $>0.87$ and is approximated as 1. The data-reduction power 
of each analysis cut is shown in Table~\ref{tab:cuts}.

The first step in offline data analysis is a set of data quality cuts.  These are particularly important
because of the volatility of the background noise.  The adaptive thresholding algorithm used online was in some
cases found to be too slow in reacting to the changing conditions. Hence, offline re-thresholding is applied
by raising the threshold whenever the rate exceeds 60 events/min (see Figure~\ref{fig:quiet_livetime}).
  The cut selecting ``quiet'' events is applied using the threshold
values after re-thresholding. Finally, events with $\Delta t_0<$~1~ms are removed, where $\Delta t_0$ is the
difference in time stamps between the current event and the previous one.  This last cut rejects bursts of
events that occasionally occur when the threshold has not adapted sufficiently and is too low.  
The combination of all event quality cuts reduced the Run B data set from 20.2 million to 2.56 million events,
with an estimated neutrino efficiency of $> 99\%$ (counting the ``quiet'' cut as a livetime reduction, not an
efficiency reduction).

\begin{table}
\caption{\label{tab:cuts}
Event rates at different stages of online (1) and offline (2-5) data reduction.  Note that the cuts in 
1-3 are
applied to single-phone events, while the cuts in 4 are applied to coincidences between 5 phones.  The
``Events" column gives the number of events surviving a particular cut and all the preceding ones.  
Cut 4a includes both coincidence windowing and hit-phone geometry (requiring the set of 5 hit phones 
to be one of the 6 trapezoidal configurations). Finally Cut 5 requires the sources to be in the geometrical
region consistent with neutrino showers and it eliminates all remaining events.}

\begin{tabular}{lll}
Cut      &       Description              	& 	Events     	\\
\hline
1) 	&	Online Triggers			&		        \\
	&	a) digital filter		&	64.6 M          \\
	&	b) correlated noise		&	20.2 M		\\
\hline			
2) 	&	Quality Cuts			&			\\
	&	a) offline re-thresholding	&	7.23 M		\\
	&	b) offline quiet conditions     &	2.60 M		\\
	&	c) $\Delta t_0 >$~1~ms		&	2.56 M		\\
\hline
3) 	&	Waveform Analysis		&			\\
	&	a) remove spikes		&	2.03 M		\\
	&	b) remove diamonds		&	1.96 M		\\
	&	c) $f_{e} >$~25~kHz		&	1.92 M		\\
\hline
4) 	&	Coincidence Building		&	        	\\
	&	a) coincidence			& 	948   		\\
	&	b) localization convergence	&	79		\\
\hline
5)      &       Geometrical fiducial region     &       0               \\
\end{tabular}
\end{table}

\begin{figure}[htb!!] 
\begin{center}
\mbox{\epsfig{file=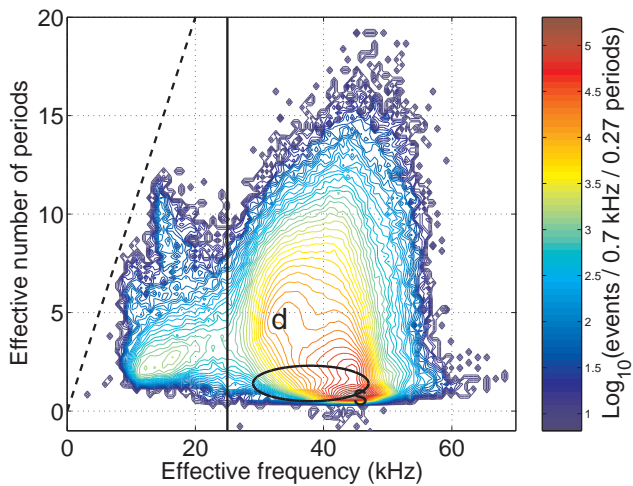,width=8.5cm}}
\caption{
Contour plot between effective frequency and effective number of periods, as defined in the text.  The dashed
line represents an upper bound due to the finite length of the capture window (1~ms). The events clustered near this
line are due to Navy pingers at specific frequencies.  The solid vertical line indicates a cut discussed in the
text.  The ellipse indicates the envelope of approximate locations of the simulated neutrino events with
different combinations of filtering and differentiation as discussed in the text.  The letters ``s'' and ``d''
identify regions where ``spike'' and ``diamond'' -type events cluster.}
\label{fig:nikolai_parameters} 
\end{center} 
\end{figure}

Waveform analysis is then used to further reduce the background.
A powerful method of event classification is provided by
the scatter plot between effective frequency $f_{e}$ and duration, measured in effective number of
cycles $n_{e}$.  We define the effective frequency as
\begin{equation}
f_e = {\frac{1}{2\pi \Delta t}} \cos^{-1}\left[{\frac{\langle P(t) P(t+\Delta t)
\rangle }{\langle P^2(t)\rangle}}\right],
\end{equation}
where $P(t)$ is the recorded pressure time series and $\Delta t$ is the sampling time
(5.6~$\mu$s), small compared to the signal oscillation period.
The effective duration in number of cycles is given by  $n_{e} = D_{e} f_{e}$, where
\begin{equation}
D_{e} = \frac {E_{S}}   {\frac{1}{2} P_{\rm peak}^2}
\label{eqn:E}
\end{equation}
is the duration of the signal in units of time. Here $P_{\rm peak}$ is the maximum absolute 
amplitude of the recorded time series and 
\begin{equation} E_{S} = E_{S+N} - E_{N} = \int P^2(t)dt - \int P_{\rm rms}^2 dt, \end{equation}
where $P_{\rm rms}$ is determined from the noise spectrum calculated once per minute.  
 The normalization is such that a 
sinusoid, in the limit of many periods in the capture window, will give $n_{e}$ equal to 
the number of cycles of the sinusoid. The contour plot between these two parameters is 
shown in Figure~\ref{fig:nikolai_parameters} for the 2.56 million events left in the data set.
Also shown in Figure~\ref{fig:nikolai_parameters} is the envelope of the regions 
containing $\sim 95$~\% of the neutrinos simulated, filtered and differentiated as described above.

The general region marked ``s'' in the contour plot includes events, here called ``spikes,''
in which a single digital sample is displaced from the rest of the waveform.  Spike events 
occurred in bursts with a 50~Hz rate and ceased when a large radar system near our DAQ
system was disabled.  We reject spike events with the metric
\begin{equation}
m_{\rm spike} = \frac{M_1(|P|)-M_2(|P|)}{M_1(|P|)}
\end{equation} 
where $P$ is the pressure time series and $M_1$ and $M_2$ give the largest and second-largest 
amplitude samples, respectively.  Events with $m_{\rm spike} < 0.7$ are retained.

\begin{figure}[htb!!]
\begin{center} 
\mbox{\epsfig{file=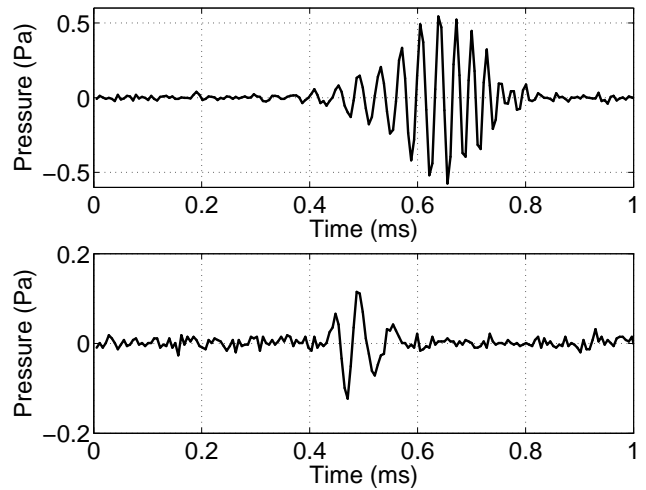,width=8.5cm}}
\caption{
Waveforms of two important classes of background.  Upper panel: ``diamond'' events.  They are easily 
rejected
with a digital matched filter.  Lower panel: ``multipolars.''  Depending on the phase response
of the SAUND detector, multipolar events could be indistinguishable from neutrino signals.}
\label{fig:background_types} 
\end{center} 
\end{figure}

The region of Figure~\ref{fig:nikolai_parameters} marked ``d'' includes events, here called
``diamonds,'' consisting of a diamond-shaped envelope containing many cycles.    These 
events,
an example of which is shown in the top panel of Figure~\ref{fig:background_types}, are believed 
to be genuine acoustic signals produced by marine mammals swimming in the area.
They can be rejected with a digital matched filter whose response function was constructed by 
averaging 10 hand-picked examples of good diamond events.  The diamond rejection metric, $m_{\rm diamond}$, is 
defined for a given waveform to be the maximum output of this digital filter acting on the waveform.
Events with $m_{\rm diamond} <0.7$
are retained.

Finally, neutrino candidates are required to have $f_{e} > 25$~kHz.  This cut eliminates low-frequency noise originating 
from sources including marine mammals, boats, and AUTEC pingers.

Waveform analysis reduces the data set from 2.56M to 1.92M events, retaining $\sim93\%$ of the
simulated neutrino events.

Coincidences between different hydrophones are reconstructed from the reduced data set. The
reconstruction program first finds cases in which 5 different phones trigger within a
2.27~s coincidence window.  This window is the maximum possible time delay between phones plus
a 10\% buffer to account for measurement errors and refraction.  We then require that the 5
events satisfy causality pairwise, a more stringent requirement than the initial windowing,
again with a 10\% buffer.  Finally we apply a phone geometry cut: There are 21 ways to choose
5 of the 7 phones.  6 of these combinations, those forming a trapezoid, are consistent with a
pancake radiation pattern triggering all phones enclosed by the pattern.  Only coincidences
triggering one of these 6 5-phone combinations are retained.

These timing and geometrical cuts greatly reduce the data set, from 1.92 million to 948 events, retaining 
an efficiency of $95\%$ for neutrino events.  Such a substantial data reduction is essential in order to 
perform the computationally intensive origin localization with TDOA in a layered medium.  
A given single-phone trigger may be included in several (even many) 5-trigger coincidences.  
This is because coincidence is a much less powerful requirement for acoustic signals than for
electromagnetic signals, because the propagation speed is $2 \times 10^5$ times lower.  While 
only four phones are required for TDOA localization, 5-phone coincidence is demanded, both to 
over-constrain the reconstruction and to reduce the background,
without significantly decreasing the neutrino efficiency.

For most coincidences, TDOA does not converge on a consistent solution to the causality
equations (solved numerically, accounting for refraction with ray-trace tables).  These
coincidences are presumably accidental and are therefore discarded.  Most of the 79 events
remaining in our data set have ``multipolar'' waveforms at more than one of the 5 phones
triggered. It is apparent that this class of events may be difficult to separate from genuine
neutrino signals on the sole basis of pulse-shape properties.  To explore the properties of
multipolar events we have built a digital matched filter using the average of 11 waveforms
from events manually chosen as good examples of this event type.  By applying the matched
filter to earlier stages of the event selection it was found that multipolar events tend
to accumulate in periods of large adaptive threshold, and many such events in the original
data set are rejected by the ``quiet conditions" (2b) cut. Clearly a full pulse-shape
calibration, including phase, of the acoustic sensors and their readout chain will be
essential for future detector arrays.  For the present data set we conservatively do {\it not}
use the multipolar event digital filter.  As will be shown in the next section, geometrical
considerations make all of these events inconsistent with a neutrino origin.

\section{Acceptance Estimate}

\begin{figure}[htb!!!!!!!!!!]
\begin{center}
\mbox{\epsfig{file=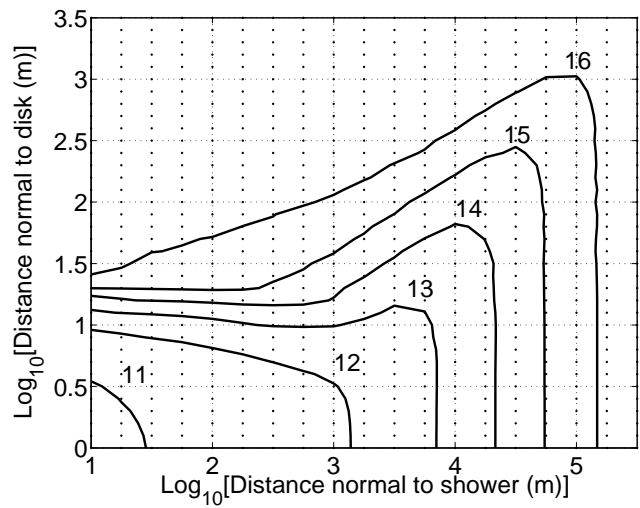,width=8.5cm}}
\caption{Neutrino detection contours for SAUND, for a trigger threshold of 0.02.  In the plot the 
neutrino producing the shower propagates along the vertical axis.  The shower maximum is at the origin 
of the coordinate system.  Curves are labeled by $\log_{10}[E ({\rm GeV})]$.  
Only one quadrant is represented.} 
\label{fig:detection_contours}
\end{center}
\end{figure}

\begin{figure}[htb!!] 
\begin{center}
\mbox{\epsfig{file=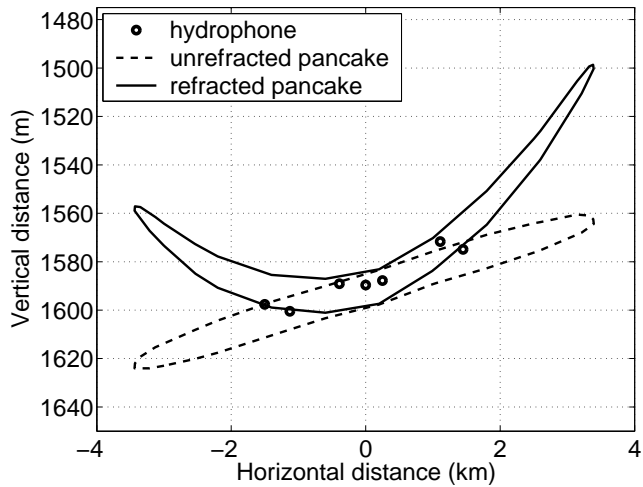,width=8.5cm}}
\caption{The effect of refraction on the acoustic radiation envelope (pancake).  A cross section 
of the acoustic radiation pattern from a $3 \times 10^{21}$~eV neutrino at $0.5^{\circ}$ zenith is
shown with and without refraction.    Projections of the seven hydrophone locations onto the cross 
sections are shown for reference.   Note that the detector is inclined to $\sim 0.5^{\circ}$ zenith, 
$7^{\circ}$ east of north, due to the slightly sloping sea floor.} 
\label{fig:refracted_pancake}
\end{center}
\end{figure}

The Monte Carlo (MC) simulation used to estimate the detector efficiency is run for a set of discrete
primary neutrino energies.  For each energy, detection contours, examples of which are given in
Figure~\ref{fig:detection_contours}, are calculated by determining the pressure pulse at each point in
space (accounting for attenuation) and then applying the digital filter.  For each energy 
$N_{\rm MC} = 10^5$ simulated neutrino events are produced with random orientations and positions in a 
water cylinder of radius 5~km, centered around the central phone.    The detection contour is then 
``bent'' to account for refraction and its intersection with the different phones in the array is
tested as shown, for a particular configuration, in Figure~\ref{fig:refracted_pancake}.
For an energy $E$ the acceptance is then given by $A_i(E) = A_0 N_i/N_{\rm MC}$,
where $N_i$ is the number of Monte Carlo events producing five or more hits at the threshold 
value identified by $i$. $A_0 = \Omega_0 V_0$ is the acceptance that would result if all $N_{\rm MC}$ events were detected, 
where $\Omega_0$ is the total solid angle, and $V_0$ the total volume within which events are chosen.


Although the outer hydrophones are only 1.5~km from the central one, it was found that events with the
appropriate orientation can safely be detected and localized up to a radial distance of 5~km.  Outside of
this region, however, the rays reach vertical turning points and precise source localization becomes
difficult.  For this reason, we conservatively limit the fiducial volume to 5~km radius.  The 
requirement
that the radiation pancake intercepts several hydrophones almost coplanar with the sea floor forces
the direction of the accepted neutrinos to be close to the vertical.

\begin{figure}[htb!!]
\begin{center}
\mbox{\epsfig{file=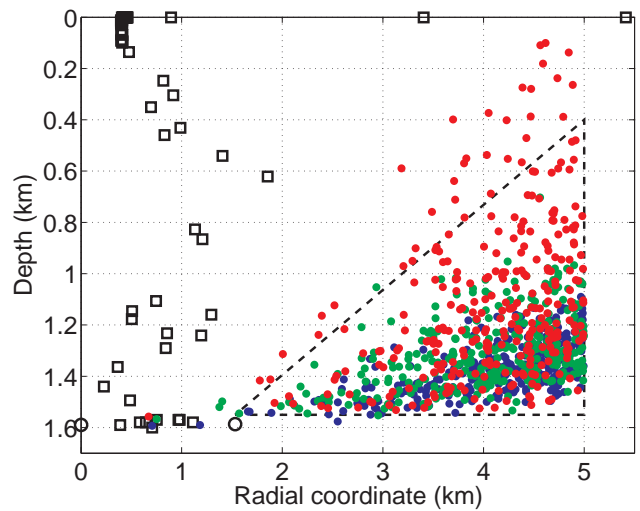,width=8.5cm}}
\caption{
Side view of reconstructed positions (squares) for the 79 coincidences surviving all cuts except for fiducial volume.  
The event with greatest radial coordinate is that at 5.4~km (no events occur beyond the plot boundary).  The
fiducial volume is bounded by the dashed triangle.  Circles represent hydrophone positions. 300 neutrino Monte Carlo 
events are also plotted, as colored dots, for each of three
different energies, $10^{14}$~GeV (blue), $10^{15}$~GeV (green), and $10^{16}$~GeV (red).}
\label{fig:event_map_side} 
\end{center} 
\end{figure}

\begin{figure}[htb!!] \begin{center}
\mbox{\epsfig{file=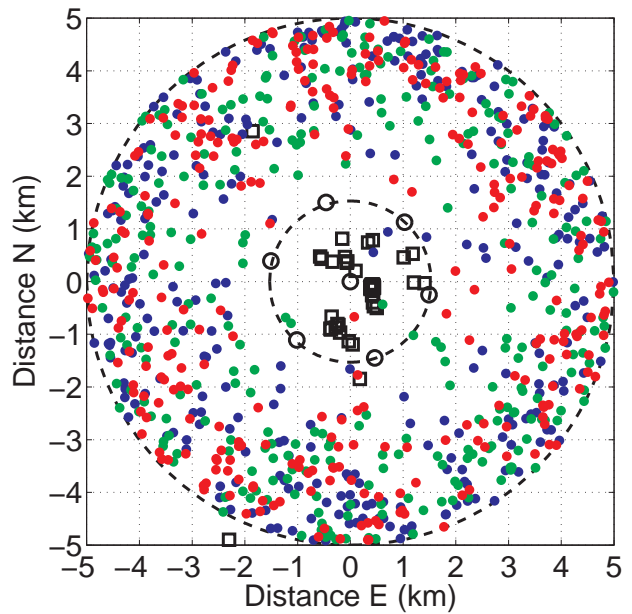,width=8.5cm}} 
\caption{Top view of the reconstructed positions (squares) for the 79 acoustic neutrino candidates    
surviving the selection cuts. The fiducial volume is bound by the two dashed circles. 
The small circles represent the positions of the hydrophones.   Neutrino Monte Carlo events for three 
different energies are also shown with the same convention as in Figure~\ref{fig:event_map_side}.} 
\label{fig:event_map_top}
\end{center}
\end{figure}

The uncertainty in source position reconstruction, estimated from Monte Carlo data, worsens steadily with distance
outside the array, but it does not exceed 500~m (vertical) and 200~m (horizontal) for radial
distances smaller than 5~km.  The
positions of detectable neutrino Monte Carlo events with energies of $10^{14}$~GeV,
$10^{15}$~GeV, and $10^{16}$~GeV are shown in a side (top) view in
Figure~\ref{fig:event_map_side} (Figure~\ref{fig:event_map_top}).  The concentration of events
slightly above the plane defined by the phones is consistent with the geometrical
considerations above.  The positions of the 79 acoustic events passing all analysis cuts (except for 
fiducial volume) in
Run B are also shown as small squares.  There is a clear separation between these events,
mainly concentrated in the water column above the hydrophones, and the region where the
neutrino events are expected to be.  Hence, we conclude that the spatial distribution of the
observed events is incompatible with the one expected from neutrino interactions. The expected
pancake shape of the acoustic emission profile from neutrinos is an essential assumption of
this assertion.    

\begin{figure}[htb!!]
\begin{center}
\mbox{\epsfig{file=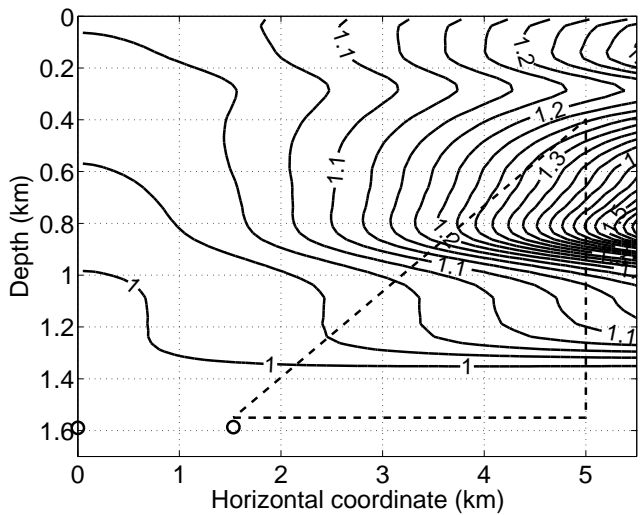,width=8.5cm}}
\caption{Degree of acoustic flux focusing at the central hydrophone, for sources at various points
as calculated from ray tracing~\citep{Boyles}.  Contours of equal focusing are given. The dashed curve
encloses the fiducial volume used for the neutrino analysis.  Central and peripheral hydrophone 
locations are indicated by circles.  Factors greater than one indicate focusing.}
\label{fig:focusing} 
\end{center}
\end{figure}

Focusing effects can potentially alter the detected energy and hence the 
threshold value.  As shown in Figure~\ref{fig:focusing}, such focusing effects are small for 
sources in the neutrino fiducial volume and are neglected.

In order to express this result as an upper limit to the ultra high energy neutrino flux, we
define a fiducial volume given by the revolution of the dashed triangle around the vertical
axis.  The fiducial region is chosen to be slightly above the ocean floor in order not to be
affected by floor roughness.  Ideally, the fiducial volume choice would have been based
entirely on the signal Monte Carlo events.  However, given the novelty of the technique and
the present emphasis on studying backgrounds, such a ``blind'' approach is impractical here.  
The separation between expected neutrino region and observed coincidences is nevertheless
striking enough that the definition of the fiducial volume is unambiguous.  In fact, the main
source of uncertainty in our analysis lies in the assumptions made about the phase response of
the array.  No 5-phone coincidences are present in SAUND Run B within the fiducial volume.

\begin{figure}[htb!!] 
\begin{center} 
\mbox{
\epsfig{file=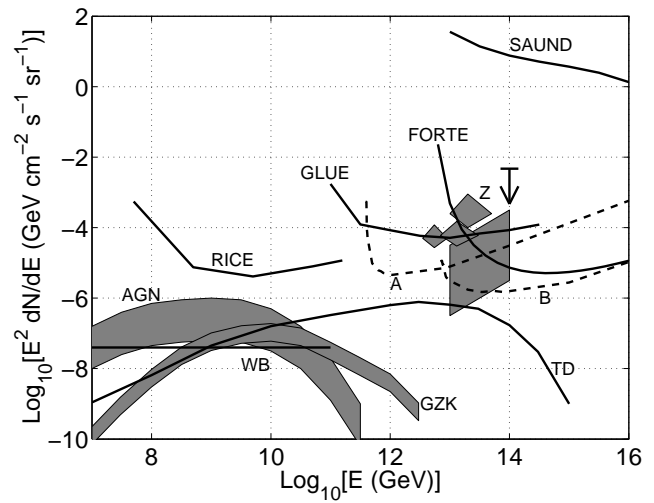,width=8.5cm}}
\caption{Diffuse neutrino flux limit from SAUND and other experiments and theoretical models of
neutrino fluxes.  The SAUND limit is the solid curve on the top-right.  The hypothetical 
sensitivities of two different optimized acoustic arrays (A and B), described in the text, 
are shown as dashed curves.   Other experimental limits (FORTE, GLUE, and RICE) are shown in 
solid black.     The remaining curves and
regions indicate various theoretical models: ``Z'' for Z-bursts (one large trapezoid, three small
quadrilaterals, and arrow), ``TD'' for topological defects, ``AGN'' for active galactic nuclei, 
``GZK'' for GZK neutrinos, and ``WB'' for the Waxmann-Bahcall limit.
 } 
\label{fig:flux_limits}
\end{center}
\end{figure}

From the geometrical acceptance $A_i(E)$ defined above, a ``neutrino exposure'' $X_i(E)=A_i(E)
N_A \sigma(E) T_i$ is computed.  Here $N_A$ is the Avogadro number, $\sigma(E)$ is the
neutrino cross section and $T_i$ is the live time for a particular value of the adaptive
threshold, labeled by $i$.  $\sigma$ is estimated by extrapolating the power-law fit given for energies
between 10$^{16}$~eV and 10$^{21}$~eV in~\citep{Gandhi}.  Fits for neutrinos are used, the
anti-neutrino fits being very similar.  Exposures collected at different thresholds are then
added together. A 90\% CL flux limit is then calculated following the method of Ref.~\citep{FORTE} 
using Poisson's
statistics and the fact that no events are observed.  The resulting limit (multiplied by $E^2$,
as customary) is shown in Figure~\ref{fig:flux_limits} along with the limits already available
from FORTE~\citep{FORTE}, RICE~\citep{RICE} and GLUE~\citep{GLUE}, and with fluxes predicted by
various theories~\citep{mod1,mod2,mod3,mod4,mod5,mod6}.

\section{Discussion}

While the main motivation of this work has been to gain operational experience and investigate
acoustic backgrounds for UHE neutrino detection, a flux measurement was obtained from the 
data.  Although the limit, shown in Figure~\ref{fig:flux_limits}, is not competitive with 
the best limits (obtained with radio techniques) it represents the first UHE
neutrino measurement performed with acoustic techniques.  A wealth of information on noise
conditions and the possible methods available to separate such noise from the neutrino signals
has been obtained.  Here we briefly discuss the characteristics of the SAUND array that turned
out to be sub-optimal for UHE neutrino detection, and we describe some of the properties that
an optimized acoustic neutrino array should have.

The importance of calibrating the phase response of the acoustic array cannot be overstated.  
Without such a calibration, it is expected that the unknown phase properties of the readout will 
dominate the systematics 
of the measurement.  An artificial source capable of providing acoustic signals similar to those 
from neutrinos with energies in the range considered here can be realized by discharging a capacitor 
in a column of sea water of length similar to that of a UHE shower.  The instantaneous heating of 
the water should closely simulate the heating produced by showers, releasing an acoustic wave with 
the correct pancake geometry and phase/frequency structure.

The flatness of the acoustic radiation lobes from UHE showers is a crucial feature that allows
noise rejection but also drastically limits the acceptance of SAUND.  Refraction exacerbates the
problem, by making regions of sea invisible to the array and by curving the emission pancake
into a shape that often has little overlap with a planar hydrophone array.  The planar
geometry of the SAUND array is also poorly optimized to provide good position resolution in the
direction orthogonal to the plane.  The vertical resolution is estimated to be as poor as 400~m for
sources deeper than 1400~m, while it is 10~m for sources close to the surface.  The resolution in
the horizontal plane, however, is better than 10~m in most of the volume.  A hydrophone array on a
3D lattice would provide substantially larger acceptance and superior pancake-shape identification.

Although the long attenuation length of sound in water allows for large (several hundred meters to several
kilometers) horizontal spacing for an optimized neutrino detector, to overcome the flatness of the
acoustic radiation an array with dense ($\sim10$~m) vertical spacing is necessary.  Even better, the ideal array
would consist not of strings of discrete modules, but of strings sensitive to acoustic signals uniformly
along their entire length.  Such a detector element is not inconceivable and could perhaps be realized
with a tube of fluid of sound speed different from the surrounding water, attached to hydrophones on
each end.  When an acoustic signal intercepts the tube, the difference in sound speeds leads to a signal
propagating in both directions along the tube.  The difference in arrival times at the two hydrophones
could be used to localize the intersection point.  Other possibilities may include interferometers built
with an optical fiber running along the length of the string (although is it not clear how this
could vertically localize signals).

To explore the possible sensitivity of an array of such strings with dense vertical spacing, 
two hypothetical arrays were tested with Monte Carlo neutrinos.  Both are hexagonal
lattices of 367 1.5~km-long strings.  The first (``A'') has 500~m nearest-neighbor spacing and is bounded
by a circle of radius $R=5$~km drawn about the central string.  The second (``B'') has 5~km spacing and is
bounded by an $R=50$~km circle.  Both are taken to have a fiducial volume given by a cylinder of height
1.5~km and radius $R+5$~km.  Circles of various radii representing acoustic radiation from neutrinos of
various energies were generated throughout this fiducial volume, with zenith from 0 to $\pi$/2.  The
calculated sensitivity of these arrays, requiring 5 strings hit, is given in
Figure~\ref{fig:flux_limits}.  

While water is the first medium to be tested for acoustic neutrino detection, salt 
(available in naturally-occurring multi-km$^3$ underground domes) and ice (available near the 
poles in multi-km deep sheets) may be more favorable.  In polar ice as compared with ocean water, 
noise is expected to be significantly lower, and signals are predicted to be greater
by an order of magnitude~\citep{Price}. It is possible that the hypothetical configurations
``A'' and ``B'' above, implemented in ice, possibly with  more advanced triggering schemes
in which thresholds are lowered for a time window after the first sensor is hit, may allow 
for better sensitivity at lower energy.   Tests are necessary in these media to determine 
their merit.  In any case acoustic detection could be employed in combination with radio and 
optical Cherenkov techniques to provide more complete information about the very rare events 
occurring at the highest energies.


\section{Acknowledgments}

We would like to thank the U.S. Navy and, in particular, D.~Belasco, J.~Cecil, D.~Deveau,
and T.~Kelly-Bissonnette, for hospitality and help at the AUTEC site.  Many
illuminating discussions with M.~Buckingham and T.~Berger (Scripps Institution of
Oceanography) are gratefully acknowledged.  We thank Y.~Zhao (Stanford) for help
with the data analysis and N.~Kurahashi for a careful reading of the manuscript.  
One of us (JV) would like to thank M.~Oreglia at the University of Chicago for 
hospitality during part of this work.  This work was supported, in part, by NSF 
Grant PHY0354497 and by a Stanford University Terman fellowship.


\end{document}